\documentclass[twocolumn,superscriptaddress,showpacs,showkeys,prl]{revtex4-1}
\usepackage{amsmath}

\newlength{\defaultparindent}
\newenvironment{annotation Text}{}{}

\begin{document}

\title{A Common Optical Algorithm for the Evaluation of Specular Spin
Polarized Neutron and M\"{o}ssbauer Reflectivities}

\author{L. De\'{a}k}
\email{deak.laszlo@wigner.mta.hu}
\affiliation{Wigner\ RCP, RMKI, P.O.B. 49, H-1525 Budapest, Hungary}
\author{L. Botty\'{a}n}
\affiliation{Wigner\ RCP, RMKI, P.O.B. 49, H-1525 Budapest, Hungary}
\author{D. L. Nagy}
\affiliation{Wigner\ RCP, RMKI, P.O.B. 49, H-1525 Budapest, Hungary}
\author{H. Spiering}
\affiliation{%
Johannes Gutenberg
Universit\"{a}t Mainz, Staudinger Weg 9, D-55099 Mainz, Germany}
\date{\today }

\begin{abstract}
Using the general approach of Lax for multiple scattering of waves a
$2\times 2$ covariant expression for the reflectivity of polarized
slow neutrons of a magnetic layer structure of arbitrary complexity
is given including polarization effects of the external magnetic
field. The present formalism is identical to the earlier published
one for the (nuclear) resonant x-ray (M\"{o}ssbauer) reflectivity and
properly takes the effect of the external magnetic field of arbitrary
direction on the neutron beam into account. The form of the
reflectivity matrix allows for an efficient numerical calculation.
\end{abstract}

\maketitle
\preprint{HEP/123-qed}

\section{Introduction}

The detectable information on a thin or stratified structure by the
reflectometric techniques is the one dimensional scattering amplitude
density profile perpendicular to the surface, which in turn can be
related to the chemical/isotopic/magnetic, etc. profile within the
penetration depth of the corresponding radiation. X-ray and neutron
reflectometry, therefore, have become standard tools in studying
surfaces and thin films. In nonresonant x-ray or unpolarized neutron
reflectometry, the scattering processes being independent of the
polarization of the incident wave, any stratified medium can be
described by a scalar complex index of refraction.  There are other
important cases, however, in which the scattering medium is
birefringent for the corresponding radiation, and the
polarization-dependent multiple scattering leads to non-scalar
optics. These cases include polarized neutron reflectometry (PNR) and
(synchrotron) M\"{o}ssbauer reflectometry (SMR), the latter being
only a special but well studied case of the anisotropic (resonant)
x-ray scattering problem. Beyond the trivial analogy between the
scalar cases of neutron and x-ray multiple scattering, the
generalization to polarization dependent scattering of {\it any} waves
\cite{Lax} is not straightforward and in fact, as we point out below,
it can not be performed in general. It is the purpose of this paper
to show that, indeed, such analogy, i.e. a common optical formalism
exists for the anisotropic neutron and anisotropic nuclear resonant
x-ray transmission and reflection for the case of forward scattering
and that of grazing incidence.

\section{General Considerations}

In this section, starting from the general theory of Lax \cite{Lax},
we shall obtain some general formulae for the scattering of
multicomponent waves. Description the theories of the various
scattering processes on a single scatterer lead to an inhomogeneous
wave equation 
\begin{equation}
\left[ \left( \Delta +k^2\right) I\ -U\left( {\bf r}\right) \right]
\Psi _1\left( {\bf r}\right) ={\bf 0,}  \label{Lax_one}
\end{equation}
where $k$ is the vacuum wave number, $I$ is the unit matrix, $U\left(
{\bf r}%
\right) $ is the scattering potential and $\Psi _1\left(
{\bf r}\right) $ is the amplitude of the scattered wave, an
electromagnetic field vector or quantum mechanical spinor state. For
many scattering centers the coherent field fulfils the
 
\begin{equation}
\left[ \left( \Delta +k^2\right) I+4\pi N\,f\right] \Psi \left( {\bf
r}%
\right) ={\bf 0,}  \label{Lax_3}
\end{equation}
three dimensional wave equation, where $f$ is the coherent forward
scattering amplitude, $N$ is the density of the scattering centers per
unit volume and $\Psi \left( {\bf r}\right) $ is the coherent field
defined by an average of the field vectors over the positions and
states of the scattering centers \cite{Lax}. Eq. (\ref{Lax_3}) shows
that from the point of view of the coherent field the system of
randomly distributed scattering centers can be replaced by a
homogeneous medium, with an index of refraction $n=I+\frac{%
2\pi
N}{k^2}f$. Since $n$ for both x-rays and slow neutrons hardly differs
from $I$, it is better to use the susceptibility tensor defined by
$\chi =$ $%
\frac{4\pi N}{k^2}f$ \cite{Deak96a}.

By choosing a simple homogeneous layer with the above susceptibility
$\chi $ and $z$ axis normal to the layer, one gets the well known 1D
wave equation: 
\begin{equation}
\Psi ^{\prime \prime }\left( z\right) +k^2\sin \theta \left[ I\sin
\theta +%
\frac \chi {\sin \theta }\right] \Psi \left( z\right) ={\bf 0.}
\label{base1}
\end{equation}
with $\theta $ being the angle of incidence. Defining $\Phi $ via
$\left( ik\sin \theta \right) \,\Phi ^{\prime }\left( z\right) :=\Psi
^{\prime \prime }\left( z\right) $, we get a system of first order
differential equations: 
\begin{equation}
\frac{{\rm d}}{{\rm d}z}\left( 
\begin{array}{c}
\Phi \\ 
\Psi
\end{array}
\right) =ikM\left( 
\begin{array}{c}
\Phi \\ 
\Psi
\end{array}
\right) ,  \label{Neuform}
\end{equation}
where 
\begin{equation}
M=\left( 
\begin{array}{cc}
0 & I\sin \theta +\frac \chi {\sin \theta } \\ 
I\sin \theta & 0
\end{array}
\right)  \label{basicM}
\end{equation}
is commonly called the ''differential propagation matrix'' in optics
\cite {Deak96a,Borzdov}. Eq. (\ref{Neuform}) was derived without
specifying the scattering process.

For an arbitrary multilayered film with homogenous layers of
thicknesses $%
d_1,d_2...d_S$ and differential propagation matrices $M_1,M_2...M_S,$
$\chi $
in Eq. (\ref{basicM}) is replaced by the susceptibility $\chi _l$ of
layer $%
l $. The solution of the differential equation (\ref{Neuform}) can be
expressed in terms of the total characteristic matrix 
\begin{equation}
L=L_S\cdot ...\cdot L_2\cdot L_1  \label{Ltotal}
\end{equation}
of the multilayer, where 
\begin{equation}
L_l=\exp \left( ikd_lM_l\right)  \label{basicL}
\end{equation}
is the characteristic matrix of the $l^{\text{th}}$ individual layer.
The $%
2\times 2$ reflectivity matrix $R$ is derived from the total
characteristic matrix $L$ by 
\begin{eqnarray}
R &=&\left( L_{\left[ 11\right] }-L_{\left[ 12\right] }-L_{\left[
21\right] }+L_{\left[ 22\right] }\right) ^{-1}  \nonumber \\ 
& & \qquad \times
\left( L_{\left[ 11\right] }+L_{\left[ 12\right] }-L_{\left[
21\right] }-L_{\left[ 22\right] }\right) ,  \label{rDeak}
\end{eqnarray}
where $L_{\left[ ij\right] }$ $\left( i,j=1,2\right) $ are $2\times
2$ blocks of the $4\times 4$ total characteristic matrix $L$
\cite{Deak96a}.  The reflected intensity 
\begin{equation}
I^r={\rm Tr}\left( R^{\dagger }R\rho \right)  \label{ReflInt}
\end{equation}
can be calculated by using the arbitrary polarization density matrix
$\rho $ of the incident beam and the reflectivity matrix
\cite{Blume}.

\section{Numerical\ Considerations}

The numerical problem in evaluating the reflectivity is the
calculation of the exponential of the $4\times 4$ matrices in Eq.
(\ref{basicL}). Here we cite our previous results \cite{EFFINO00}
proving that it is possible to get a closed solution to the general
problem requiring the calculation of $%
2\times 2$ matrices only. The characteristic matrix is of the form
\begin{equation}
L_l=\left( 
\begin{array}{cc}
\cosh \left( kd_lF_l\right) & \frac 1xF_l\sinh \left( kd_lF_l\right)
\\ xF_l^{-1}\sinh \left( kd_lF_l\right) & \cosh \left( kd_lF_l\right)
\end{array}
\right) ,  \label{LDeak}
\end{equation}
where the $2\times 2$ $F_l=\sqrt{-I\sin ^2\theta -\chi _l}$ and
$x=i\sin
\theta $ \cite{Deak96a}.

To evaluate Eq. (\ref{LDeak}), first we have to calculate the $2\times
2$ square root of the $F$ matrices. This can be made by using the
identity 
\begin{equation}
G^{1/2}=\frac{G+I\sqrt{\det G}}{\sqrt{%
\mathop{\rm Tr}
G+2\sqrt{\det G}}},  \label{sqr22}
\end{equation}
where $G$ is any nondiagonal $2\times 2$ matrix \cite{Borzdov}. The
$\sinh $ and $\cosh $ functions are calculated from their definition
with the exponential functions. Moreover, the exponential of the
$2\times 2$ matrix $%
G $ can be expressed by itself and its scalar invariants: 
\begin{eqnarray}
\exp G &=& \exp (\frac 12%
\mathop{\rm Tr}
G)  \notag \\
& & \times \left[ \cos \sqrt{\det \bar{G}}I+\frac{\sin \sqrt{\det
\bar{G}}}{\sqrt{%
\det \bar{G}}}\bar{G}\right] ,  \label{exp22}
\end{eqnarray}
where $\bar{G}=G-%
{\textstyle {1 \over 2}}
I%
\mathop{\rm Tr}
G$ \cite{Blume}.

In order to calculate the characteristic matrix of a semi-infinite
layer (substrate) $S$, we have to find its $L_S\rightarrow L^\infty $
limit for $%
d_S\rightarrow \infty $. From Eqs. (\ref{basicM}) through
(\ref{exp22}) follows that the corresponding limit is given by 
\begin{equation}
L^\infty =\left( 
\begin{array}{cc}
I & p\sqrt{I+\frac{\chi _S}{\sin ^2\theta }} \\ 
p\left( \sqrt{I+\frac{\chi _S}{\sin ^2\theta }}\right) ^{-1} & I
\end{array}
\right)  \label{Linfinit}
\end{equation}
where $p=%
\mathop{\rm sgn}
\left[ 
\mathop{\rm Re}
\left( 
\mathop{\rm Tr}
F_S\right) \right] $ is the sign of the real part of the trace of
$F_S.$

The above algebra turns out to be numerically very stable, therefore
this approach is suitable for fast numerical calculations of the
characteristic matrices for anisotropic stratified media. In fact,
the exponential of the matrix in Eq. (\ref{basicM}) can be calculated
exactly without solving any eigenvalue problem. The program based on
this calculus is freely available \cite{EFFINO00,EFFINOweb}.

\section{M\"{o}ssbauer and Polarized Neutron Reflectometries}

A simple application of Eq. (\ref{Neuform}) to nuclear resonant x-ray
scattering is not possible, since the anisotropic Maxwell equations
and the spin-dependent Schr\"{o}dinger-equation lead to different
results \cite {Borzdov,Andreeva1} and the $3\times 3$ susceptibility
tensor can not be expressed by the $2\times 2$ forward scattering
amplitude $f$ in general.  However, starting from the Maxwell
equations and using the $3\times 3$\ nuclear susceptibility tensor
given by Afanas'ev and Kagan \cite{Kagan} the nuclear resonant x-ray
reflectivity could be derived \cite{Deak96a} for forward scattering
and grazing incidence in terms of the coherent forward scattering
amplitude. The dynamical theory of x-ray scattering \cite
{Hannon4,Roehlsberger} provide an equivalent result in the grazing
incidence limit \cite{Deak96a,EFFINO00}{\em .} However, in
\cite{Deak96a} both an upper and a lower limit was found for the
grazing angle $\theta $ for this approximation to apply, which limits
are not present in the original theory of Lax \cite{Lax}. The forward
scattering amplitude matrix was expressed for the nuclear resonant
x-ray case in \cite{Blume,Spiering85} in terms of the hyperfine
interactions.

The application of the above optics for PNR implies specifying $f$ (or
$\chi $) for the interaction potential $U$ in Eq. (\ref{base1}). We
use the potential $U\left( {\bf r}\right) =U_p\left( {\bf r}\right)
+U_m\left( {\bf r%
}\right) $ as the sum of the isotropic nuclear potential 
\begin{equation}
U_p\left( {\bf r}\right) =4\pi b\delta \left( {\bf r}\right) I,
\label{Ppotential}
\end{equation}
and the anisotropic magnetic potential 
\begin{eqnarray}
U_m\left( {\bf r}\right) &=&-\frac{2m}{\hbar ^2}{\bf \mu }_m\cdot \left[
{\bf B%
}_a\left( {\bf r}\right) +{\bf B}_{ext}\right] \nonumber \\
&=&-\frac{2m}{\hbar
^2}{\bf \mu }_m\cdot {\bf B}\left( {\bf r}\right)  \label{Mpotential}
\end{eqnarray}
with $m$ being the mass of the neutron, $b$ the nuclear scattering
length of the nucleus in the laboratory system, ${\bf \mu }_m=g\mu
_N{\bf \sigma }$ the magnetic moment operator of the neutron,
$g=-1.9132$, $\mu _N{\bf =}%
5.050\times 10^{-27}$Am$^2$, ${\bf \sigma }$ the Pauli operator, ${\bf
B}_a$ the atomic magnetic field, ${\bf B}_{ext}$ the (homogeneous)
external magnetic filed. In the first Born approximation 
\begin{equation}
f=-\frac 1{4\pi }\int\limits_\Omega d^3{\bf r\ }U\left( {\bf r}\right)
,
\label{fnull}
\end{equation}
where $\Omega $ is the volume of the interaction (in fact the atomic
volume). By using $\chi =\frac{4\pi N}{k^2}f$ we get 
\begin{equation}
\chi =\frac 1{k^2}\left[ \frac{2m}{\hbar ^2}g\mu _N{\bf \sigma \cdot
}%
\overline{{\bf B}}-4\pi N\,\sum_i\alpha _ib_iI\right] , 
\label{khiNeutron}
\end{equation}
where index $i$ accounts for the different types of scattering
centers, and $%
\alpha _i$ for the relative abundance of the $i$th nucleus. The mean
magnetic field $\overline{{\bf B}}={\bf B}_{ext}+\overline{{\bf
B}}_a={\bf B}%
_{ext}+\frac 1\Omega \int\limits_\Omega d^3{\bf r\ B}_a\left( {\bf
r}\right) .$

In neutron reflectometry the scattering vector, $Q=2k\sin \theta $ and
the scattering length density 
\begin{equation}
K=k^2\chi =\frac{2m}{\hbar ^2}g\mu _N{\bf \sigma \cdot }\overline{{\bf
B}}%
-4\pi N\,\sum_i\alpha _ib_iI  \label{Kneutron}
\end{equation}
are more often used than $\theta $ and $\chi $. With these notations
Eq. (%
\ref{Neuform}) reads 
\begin{equation}
\frac{{\rm d}}{{\rm d}z}\left( 
\begin{array}{c}
\Phi \\ 
\Psi
\end{array}
\right) =i\left( 
\begin{array}{cc}
0 & \frac Q2I+\frac{2K}Q \\ 
\frac Q2I & 0
\end{array}
\right) \left( 
\begin{array}{c}
\Phi \\ 
\Psi
\end{array}
\right) .  \label{NNeuform}
\end{equation}
Using the definition of the Pauli matrices, the scattering length
density matrix Eq. (\ref{Kneutron}) is expressed by the physical
quantities 
\begin{eqnarray}
K &=& \frac{2m}{\hbar ^2}g\mu _N\left( 
\begin{array}{cc}
\overline{B}_x & \overline{B}_y-i\overline{B}_z \\ 
\overline{B}_y+i\overline{B}_z & -\overline{B}_x
\end{array}
\right) \nonumber \\ 
&-&4\pi N\,\sum_i\alpha _ib_iI,  \label{KBvel}
\end{eqnarray}
where $\overline{B}_x,\overline{B}_y,\overline{B}_z$ are the
components of the magnetic field $\overline{{\bf B}}$.

Having $K$ from Eqs. (\ref{KBvel}) and (\ref{Kneutron}) for each layer
$l$, Eq. (\ref{exp22}) is used to calculate the exponential of the
differential propagation matrix of Eq. (\ref{NNeuform}) is obtained.
With this (by applying (\ref{LDeak}) to (\ref{Linfinit})) first the
(\ref{basicL}) characteristic matrices , then the (\ref{Ltotal})
total characteristic matrix $L$, from which the (\ref{rDeak}) complex
reflectivity matrix $R$ is calculated. For the sake of brevity, we
dropped the layer index $l$ in $K,$ $%
\overline{{\bf B}}$, $N$, $\alpha _i$ and $b_i$ in Eqs. (\ref{fnull})
to (%
\ref{KBvel}).

An elegant covariant treatment of specular PNR \cite{Broki99}
including earlier matrix methods of restricted form
\cite{Felcher87,Majkrzak89} recently published by R\"{u}hm et. al.
turns out to be equivalent to the present results. Indeed,
substituting $p_0=k\sin \theta $\ and $\widehat{%
{\cal H}}_l=-\left( \hbar ^2k^2/2m\right) \chi _l$\ for layer $l$\ in
Eq.  (7) of \cite{Broki99} we obtain (\ref{LDeak}), an equation
equivalent to Eq.  (3.20) of \cite{Deak96a}. Consequently, what we
have shown here is the equivalence \cite{DeakPhD} of the supermatrix
formalisms developed for SMR \cite{Deak96a} and PNR \cite{Broki99}.

\section{The External Magnetic Field as an Anisotropic Medium}

Although their general treatment would have allowed for, R\"{u}hm et
al. \cite{Broki99} did not explicitely studied the effect of the
(guiding or polarizing) external magnetic field on the neutron beam,
what we briefly outline in this section in the standard manner
borrowed from anisotropic optics \cite{Borzdov}.

The (\ref{rDeak}) reflectivity expression is only valid for a neutron
beam incident on the layer system ($l=1,2,3,..,S$) from the vacuum
($l=0$). In the typical experimental setup, however, guiding fields
and often strong external magnetic fields are used in order to
eliminate depolarization of the neutrons and to ensure polarization
of the sample, respectively. The effect of the external magnetic
field was studied by Pleshanov \cite {Pleshanov} and Fermon
\cite{Fermon} in detail. From Eq. (\ref{KBvel}) it follows, that the
vacuum, in presence of an external magnetic field, is an anisotropic
'medium'. Consequently, the incoming beam is given in this 'medium'
instead of being given in the vacuum. In order to treat this problem,
following Borzdov \cite{Borzdov} (for a brief outline in English see
\cite{Deak96a}), for the case of neutron reflectometry we introduce
an impedance tensor $\gamma $ by the following relationship: 
\begin{equation}
\gamma ^{0,r,t}\Psi ^{0,r,t}:=\Phi ^{0,r,t},  \label{gammaNeu}
\end{equation}
where indexes $0,r$ and $t$ indicate incident, reflected and
refracted waves, respectively (see Eq. (3.4) of Ref. \cite{Deak96a}).
Substituting (%
\ref{gammaNeu}) into Eq.(\ref{NNeuform}) we get the impedance tensors
\begin{equation}
\gamma \equiv \gamma ^0=-\gamma ^r=\sqrt{I+\frac{4K}{Q^2}}, 
\label{GammaKif}
\end{equation}
where $K$ is calculated from Eq. (\ref{KBvel}) for the given external
magnetic field. We dropped $\gamma ^t$ because the substrate is taken
as a semi-infinite layer with Eq. (\ref{Linfinit}). From expressions
(\ref {Kneutron}) and (\ref{KBvel}) for $\gamma $ we get 
\begin{equation}
\gamma \left( Q\right) =\frac 1Q\left( 
\begin{array}{cc}
\sqrt{Q_{+}} & 0 \\ 
0 & \sqrt{Q_{-}}
\end{array}
\right) ,  \label{GammaEig}
\end{equation}
where 
\begin{equation}
Q_{\pm }^2=Q^2\pm \frac{8m}{\hbar ^2}g\mu _N\left| {\bf
B}_{ext}\right|
\label{Q0Q}
\end{equation}
is the momentum $Q_{\pm }=2k\sin \theta _{\pm }$ measured in the
external magnetic field. Due to the birefringence of anisotropic
media (including vacuum in presence of external magnetic field), the
beam propagation directions for the different polarizations
necessarily differ from each other, consequently the angles of
incidence and the momentum of the beams with different polarizations
(sign '$+$' and '$-$') are also different ($%
\theta _{\pm }$ and $Q_{\pm }$). The vacuum momentum $Q$ can be
calculated backwards from $Q_{\pm }^2$ by applying the Fresnel
refraction law [Eqs. (%
\ref{KBvel}) and (\ref{base1}), as well as the definition of $K$ and
$Q$ by Eq. (\ref{Kneutron})].

Having the impedance tensors of the individual layers, we simply apply
the modified general $2\times 2$ reflectivity expression 
\begin{eqnarray}
R &=&\left[ \left( L_{\left[ 11\right] }-L_{\left[ 21\right] }\right)
\gamma -L_{\left[ 12\right] }+L_{\left[ 22\right] }\right] ^{-1} 
\nonumber \\ 
& & \qquad \times \left[ \left( L_{\left[ 11\right] }-L_{\left[
21\right] }\right) \gamma +L_{\left[ 12\right] }-L_{\left[ 22\right]
}\right]  \label{RfinNeu}
\end{eqnarray}
which takes the effect of the external magnetic field into account
through the impedance tensor $\gamma $ \cite{Deak96a}. The reflected
intensity $I^r$ is calculated from Eq. (\ref{ReflInt}) using the the
reflectivity matrix $R$ and the polarization density matrix $\rho =$
$\overline{\left| \Psi \right\rangle \left\langle \Psi \right| }$ of
the incident beam, where the bar represents the average over the
polarizations \cite{Blume,Broki99}.

\section{ CONCLUSION}

In summary, a common optical formalism of (nuclear) resonant x-ray
(M\"{o}ssbauer) reflectometry and polarized neutron reflectometry was
presented. Consequently, the strictly covariant formalism of
\cite{Borzdov} as published in \cite{Deak96a} and the corresponding
computer program \cite {EFFINO00,EFFINOweb} are readily available for
neutron reflectometry of layered systems of arbitrary complexity.
Taking the effect of the external magnetic field through the
impedance tensor into account, a modified reflectivity expression is
given.{\em \ }The form of the reflectivity matrix allows for a very
efficient numerical algorithm for both SMR and PNR implemented in
\cite{EFFINO00,EFFINOweb}.

 \begin{acknowledgments}
This work was partly supported by the Hungarian Scientific Research
Fund (OTKA) under Contract Nos. T029409 and F022150. L. D. thanks for
support by the Deutscher Akademischer Austauschdienst (DAAD).
 \end{acknowledgments}

\end{document}